\documentclass[aps,prd,twocolumn,superscriptaddress,nofootinbib]{revtex4-2}

\usepackage[colorlinks=true,citecolor=red,urlcolor=blue]{hyperref}
\usepackage{amssymb}
\usepackage{amsmath,color}
\usepackage{graphicx}
\usepackage{bbm}
\usepackage{verbatim}
\usepackage[T1]{fontenc}
\usepackage[utf8]{inputenc}
\usepackage[normalem]{ulem}

\usepackage{url}
\usepackage{xcolor}
\usepackage{caption}
\usepackage{subcaption}

\usepackage{diagbox}
\usepackage{color}
\usepackage{multirow}
\usepackage{hhline}
\usepackage{tabularx}

\newcommand{\orcid}[1]{\href{https://orcid.org/#1}{\hspace{0.5mm}\raisebox{-0.ex}{\includegraphics[height=2.0ex]{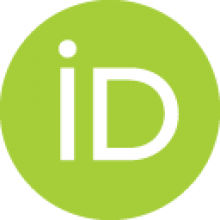}}}}

\begin{document}

\title{Dynamical Dark Energy Implies a Coupled Dark Sector: Insights from DESI DR2 via a Data-Driven Approach}


\author{Changyu You}
\affiliation{School of Physics and Technology, Wuhan University, Wuhan 430072, China}  

\author{Dan Wang}
\affiliation{School of Physics and Technology, Wuhan University, Wuhan 430072, China}

\author{Tao Yang\orcid{0000-0002-2161-0495}}
\email[Corresponding author ]{yangtao@whu.edu.cn}
\affiliation{School of Physics and Technology, Wuhan University, Wuhan 430072, China}

\date{\today}

\begin{abstract}
Recent observations from the Dark Energy Spectroscopic Instrument (DESI) Data Release 2 (DR2) have revealed compelling evidence for dynamical dark energy, challenging the $\Lambda$CDM paradigm. In this work, we adopt a data-driven, model-independent approach to reconstruct the dark energy equation of state (EoS) and its potential interaction with dark matter using combined background cosmological datasets, including DESI DR2, cosmic chronometers, observational Hubble data, and Type Ia supernovae. Using Gaussian Process regression and a non-parametric formalism, we first confirm a $\sim 2\sigma$ indication of dynamical dark energy, featuring a phantom crossing around redshift $z \sim 0.4$, consistent with DESI results. We then explore the implications of dynamical EoS from DESI DR2 for dark sector coupling. Incorporating priors on the EoS from DESI DR2, we find a $\sim 2\sigma$ signal for non-zero interactions between dark energy and dark matter at low redshift. Our results suggest that if DESI’s evidence for time-varying dark energy is confirmed, a coupled dark sector may be a necessary extension beyond $\Lambda$CDM.
\end{abstract}

\maketitle

\section{Introduction}
The $\Lambda$CDM cosmology, established as the standard paradigm in modern cosmology, 
has been highly successful in explaining a wide range of astrophysical phenomena --
from cosmic microwave background (CMB) anisotropies to large-scale structure formation
\citep{SupernovaSearchTeam:1998fmf,SupernovaCosmologyProject:1998vns,WMAP:2012nax,Planck:2018vyg,SDSS:2003eyi,SDSS:2005xqv}.
However, growing discrepancies between early- and late-universe observations -- 
most notably the Hubble tension (a $\sim 5\sigma$ discrepancy in $H_0$~\citep{Riess:2021jrx,Planck:2018vyg}), 
the $S_8$ tension~\citep{Planck:2018vyg,DES:2025xii}, 
and cosmic shear anomalies~\citep{Mccarthy:2017yqf,Hildebrandt:2018yau,Asgari:2019fkq} -- challenge the model’s validity. 
Proposed resolutions include modifications to gravity~\citep{Kunz:2006ca,Clifton:2011jh,Hu:2007nk} 
and dynamical dark energy scenarios~\citep{Copeland:2006wr,Tsujikawa:2013fta,Armendariz-Picon:2000ulo}.

The first data release (DR1) of the Dark Energy Spectroscopic Instrument (DESI)~\citep{DESI:2024mwx} suggested evidence for dynamical dark energy, 
inconsistent with the $\Lambda$CDM framework. 
This result has spurred significant interest in constraining the dark energy equation of state (EoS) through 
various data analysis approaches~\citep{DESI:2024kob,Shlivko:2024llw,Tada:2024znt,Giare:2024gpk,Gialamas:2024lyw}. 
More recently, the second data release (DR2) from DESI~\citep{DESI:2025zgx}, based on three years of observations, has reinforced these findings. 
Using a broad range of parametric and non-parametric methods, such as the Chevallier-Polarski-Linder (CPL) parameterization~\citep{Linder:2002et,Planck:2018vyg,Planck:2015bue,Pan-STARRS1:2017jku} and Gaussian Processes,
DESI DR2 reports a 2.8$\sigma$ to 4.2$\sigma$ statistical preference for time-varying dark energy over the standard $\Lambda$CDM model~\citep{DESI:2025zgx,DESI:2025fii,DESI:2025wyn}.

The study of dynamical dark energy models has intensified in recent decades, 
with each model offering distinct advantages and limitations~\citep{Copeland:2006wr,Armendariz-Picon:2000ulo,Linder:2007wa,Pan-STARRS1:2017jku}.
Quintessence -- a widely studied dynamical model in which dark energy is represented by a scalar field -- 
has received significant attention~\citep{Tsujikawa:2013fta,Linder:2007wa,Saridakis:2010mf}. 
However, recent DESI DR2 results challenge this framework by revealing a phantom crossing at $z \simeq 0.4$, 
with $w > -1$ at $z < 0.4$ and $w < -1$ at $z > 0.4$, a behavior not permitted in standard quintessence models.
Among the various parameterizations designed to capture such dynamics, 
the CPL form, $w(z) = w_0 + w_a z / (1 + z)$, 
stands out as the most widely used phenomenological approach and has been extensively constrained with DESI BAO data recently~\citep{DESI:2024mwx,DESI:2025zgx,Wang:2024rjd,Chan-GyungPark:2024brx}. 
In this work, we model dark energy as a phenomenological fluid, reconstruct its EoS using a model-independent (non-parametric) approach, 
and reassess its evolution in light of the DESI DR2 results.

Another scenario beyond the $\Lambda$CDM model involves considering interactions between the dark sectors. 
Given the unknown nature of both dark matter (DM) and dark energy (DE), it is plausible that they may interact through non-zero couplings.
This possibility has motivated the development of interacting dark energy (IDE) models and its corresponding methodology
\citep{Amendola:1999er,Wang:2016lxa,Cai:2004dk,Boehmer:2008av,Amendola:2003eq,DiValentino:2019ffd,Yang:2018euj,Zhang:2018glx,Liu:2022hpz,Zhao:2022ycr,Nunes:2022bhn}, 
which may help alleviate the aforementioned cosmological tensions and the coincidence problem~\citep{Weinberg:2000yb,DiValentino:2019jae,Sabogal:2025mkp,Sabogal:2024yha,Benisty:2024lmj,DiValentino:2017iww}. 
The interaction between dark energy and dark matter has been examined using recent DESI data, based on a specific form of the coupling function~\citep{Giare:2024smz,Li:2024qso,Chakraborty:2025syu}. 
Notably, \citet{Giare:2024smz} reported a statistically significant interaction at the $95\%$ confidence level with DESI DR1, suggesting an energy-momentum transfer from dark matter to dark energy. 
However, given the lack of a theoretically favored interaction model, it is more appropriate to constrain the dark sector coupling using a model-independent approach.

In this paper, we employ a non-parametric, data-driven method~\citep{Yang:2015tzc,Yang:2020jze} to reconstruct the interaction between DE and DM using background cosmological data sets, including DESI DR2.
We highlight the degeneracy between dark sector coupling and the EoS of dark energy in the context of this model-independent framework. 
By incorporating priors on the dynamical EoS from the DESI DR2 results, we identify approximately $2\sigma$ evidence for non-zero dark sector interactions at low redshift, with energy transferred from dark matter to dark energy.  
Our result is consistent with findings from the parametric approach~\cite{Giare:2024smz}. 
Our analysis indicates that, from a non-parametric, data-driven perspective using current background cosmological data sets, the dynamical dark energy inferred from DESI DR2 inevitably implies a coupled dark sector.
This conclusion is independent of any specific cosmological model.



\section{Method and data}
In interacting dark sector scenarios, the dark matter density evolves as $\rho_c \sim f/a^3$, rather than following the standard $a^{-3}$ scaling, 
where $f$ is an arbitrary function representing the coupling between dark energy and dark matter~\citep{Das:2005yj}.
To remain consistent with the equivalence principle and Solar System tests of gravity~\citep{Bertotti:2003rm,Will:2014bqa}, 
we do not couple dark energy to baryons~\footnote{Introducing a universal coupling between dark energy and the total non-relativistic matter (including subdominant baryons) does not alter our conclusions.}. Within the flat Friedmann-Lemaître-Robertson-Walker(FLRW) framework, the Friedmann equations and continuity equations for every species take the form:

\begin{eqnarray} 
&3H^2=\rho_{de}+\rho_c+\rho_b\,, \label{eq:Friedmann}   \\
&\dot\rho_b+3H\rho_b=0    \label{eq:rhob}\,, \\
&\dot\rho_c+3H\rho_c=3H_0^2\Omega_c\frac{\dot f}{a^3f_0} \,,\\
&\dot\rho_{de}+3H(1+w_{de})\rho_{de}=-3H_0^2\Omega_c\frac{\dot f}{a^3f_0} 
\end{eqnarray}

Here, $a = 1/(1+z)$ is the scale factor, $H = \dot{a}/a$ is the Hubble parameter, 
$f_0$ denotes the present-day value of the coupling function $f$, and dots represent derivatives with respect to time. 
The density parameter for a component $X$ is given by $\Omega_X = \rho_{X0} / (3H_0^2)$, where $\rho_{X0}$ is its present energy density.
Since we focus on the late-time Universe, where dark energy and matter dominate, the radiation contributions are ignored.  
For simplicity, we adopt natural units with $M_{\rm pl}=1$ throughout our analysis.

To avoid specifying prior information on the three parameters $f_0$, $\Omega_c$, and $\Omega_b$ during the reconstruction of the coupling, 
we use $\Omega_f = \Omega_c \frac{f}{f_0} + \Omega_b$ to investigate the interaction between dark energy and dark matter. 
A constant $\Omega_f$ (i.e., $\dot\Omega_f=0$) indicates the absence of coupling between the dark sectors. 
Using Eqs.~(\ref{eq:Friedmann}-\ref{eq:rhob}) we get the EoS of dark energy: 
\begin{equation}
    w_{de}=\frac{(-2\dot H-3H^2)}{(3H^2-3H_0^2\frac{\Omega_f}{a^3})}\label{eq:w_phi}
\end{equation}

In the uncoupled limit ($f\equiv f_0$), $\Omega_f$ reduces to $\Omega_c+\Omega_b=\Omega_m$. 
The equation above enables the reconstruction of the EoS of dark energy directly from cosmological expansion data.

For coupled dark energy, the coupling and the EoS of dark energy are related through Eq.~(\ref{eq:w_phi}),
\begin{eqnarray}
\Omega_f=\frac{a^3}{3H_0^2} && \left(3H^2+\frac{2\dot H+3H^2}{w_{de}}\right)\,, \label{eq:f}\\
\dot \Omega_f=\frac{a^3}{3H_0^2} && \left((9H^3+6H\dot H)+ \frac{2\ddot H+12H\dot H+9H^3}{w_{de}} \right .\nonumber \\
&& \quad\left .-\frac{(2\dot H+3H^2)\dot w_{de}}{w_{de}^2}\right) \,. 
\label{eq:dotf}
\end{eqnarray}

The dark energy EoS $w_{de}$ exhibits degeneracy with dark sector interactions in background expansion observables (e.g., Hubble parameter, luminosity distances). 
For a fixed expansion history, 
reconstructing either EoS ($w_{de}$) or coupling ($\Omega_f$ or $\dot{\Omega}_f$) requires prior knowledge of the other component. 
This degeneracy implies that dynamical dark energy and dark sector couplings can mutually compensate for each other, 
yielding identical expansion signatures in observational data.

To reconstruct these parameters from observational data, 
we convert all time derivatives to redshift derivatives using the transformation  $\frac{d}{dt}=-H(1+z)\frac{d}{dz}$. 
Given a prior on $w_{de}$, our framework can simultaneously reconstruct the coupling parameters: $\Omega_f$ and its evolution $\dot{\Omega}_f$. A non-zero $\dot{\Omega}_f$ signals the presence of interactions between the dark sectors.

The data sets used in our analysis are listed below:

\begin{itemize}

\item 6 $H(z)$ data from DESI DR2~\citep{DESI:2025zgx}. The DESI-BAO measurements are expressed by three ratios: the three-dimensional BAO mode $\tilde{D}_V$, the transverse mode $\tilde{D}_M$,and the radial mode $\tilde{D}_H$. In this work, we focus on the radial mode.  

Along the line of sight we measure:
\begin{equation}
    \tilde{D}_H= \frac{D_H}{r_d}=\frac{c}{r_dH(z)}
\end{equation}
So we get $H(z)$ :
\begin{equation}
    H(z)=\frac{c}{r_d\frac{D_H}{r_d}}=\frac{c}{r_d\tilde{D}_H}
\end{equation}

\item 18 $H(z)$ data from the homogenized model-independent OHD (BAO features) compiled in Table 2 of~\citet{Magana:2017nfs}. We replace the three highest-redshift Lyman-$\alpha$ (Ly$\alpha$) BAO measurements~\citep{BOSS:2013igd,BOSS:2014hwf,BOSS:2017fdr} at
$z=2.33,2.34$ and 2.36 with the updated $z=2.34$ eBOSS DR14 result~\citep{eBOSS:2019qwo,eBOSS:2019ytm}.
The $H(z)$ value at this redshift is derived using the same methodology applied to DESI BAO data. 
Note that we adopt the same sound horizon $r_d=147.33$ Mpc  at the drag epoch using in~\citep{Magana:2017nfs} from Planck measurements to ensure that all $H(z)$ from BAO have the same prior on $r_d$.

\item 31 $H(z)$ data from cosmic chronometers (CC) obtained using the differential-age technique in Table 1 of~\citet{Gomez-Valent:2018hwc}. 

\item 6 $E(z)$ data from Pantheon+MCT SNe Ia Measurements given by~\citet{Riess:2017lxs}. Here $E(z)=H(z)/H_0$ denotes the dimensionless Hubble parameter. For the $z=1.5$ point, we employ a Gaussian likelihood approximation following \citet{Gomez-Valent:2018gvm}, incorporating the full covariance matrix of uncertainties.
\end{itemize}

\section{Reconstructions and results}
To ensure self-consistency across the four observational datasets employed in our reconstructions, 
we standardize all $H(z)$ data measurements to the dimensionless Hubble parameter $E(z)\equiv H(z)/H_0$, adopting a fiducial $H_0$ value. 
We use Gaussian Process (GP) regression\footnote{\url{https://github.com/JCGoran/GaPP/tree/feature/python3}}, 
a machine learning method notable for its ability to predict the evolution of derived cosmological parameters in a model-independent manner~\citep{Seikel:2012uu,Shafieloo:2012ht}.
The fiducial value of $H_0$ is determined using the combined DESI DR2+OHD+CC data. 
We then reconstruct $E(z)$, $E'(z)$, $E''(z)$, and $E'''(z)$ based on the combination of all four datasets. 
We obtain a mean value of $H_0 = 68.87~\mathrm{km\,s^{-1}\,Mpc^{-1}}$, 
which lies between the estimates derived from Planck~\citep{Planck:2018vyg} and SNe Ia~\citep{Riess:2021jrx}.
We rewrite Eqs.~(\ref{eq:w_phi}-\ref{eq:dotf}) to substitute $H(z)$ with $E(z)$ by dividing $(H_0)^n$ on both sides of the equations. 
Here, $n$ corresponds to the dimensional order of each equation with respect to $H_0$. 
Then the dimensionless parameters we can reconstruct are actually $w_{de}$, $\Omega_f$, $\frac{\dot \Omega_f}{H_0}$, etc.

\begin{figure*}
\includegraphics[width=0.45\textwidth]{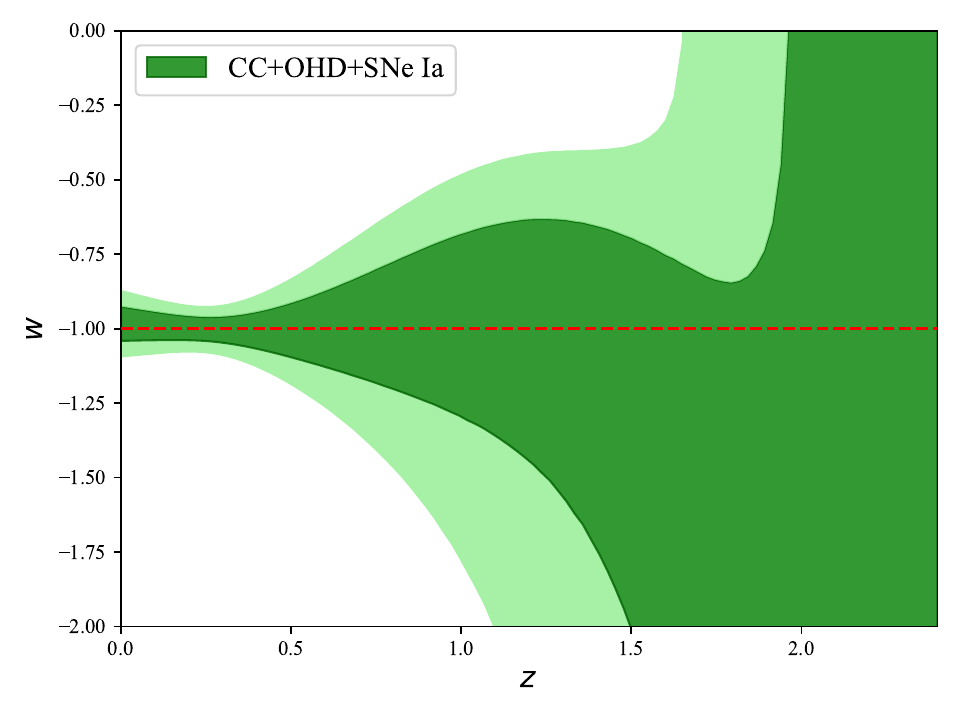} 
\includegraphics[width=0.45\textwidth]{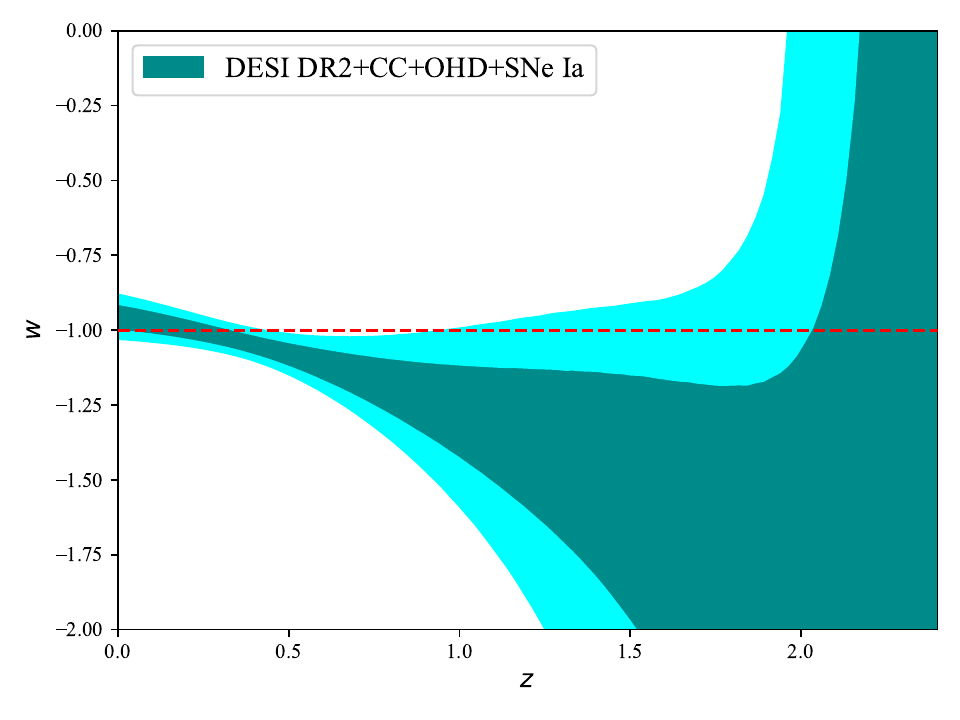}
\caption{Reconstructions of the dark energy equation of state in the uncoupled case using different datasets. We present the $68\%$ and $95\%$ confidence levels (CL) as dark and light shaded bands, respectively. Dashed lines represent the $\Lambda$CDM model ($w = -1$).}
\label{fig:uncoupled}
\end{figure*}

For the uncoupled case, we adopt the value $\Omega_f = \Omega_m = 0.31$ from Planck~\citep{Planck:2018vyg}. 
The equation of state for uncoupled dark energy, reconstructed using different combinations of data sets, is shown in Fig.~\ref{fig:uncoupled}. 
Note that for the reconstruction of $w(z)$ using the CC+OHD+SNIa data combination, 
the fiducial value of $H_0$ is $68.06~\mathrm{km\,s^{-1}\,Mpc^{-1}}$, 
derived from the same data set with GP. 
As shown in Fig.~\ref{fig:uncoupled}, without DESI BAO data, 
the reconstruction of $w_{de}$ is fully consistent with the $\Lambda$CDM model ($w = -1$). 
However, incorporating DESI DR2 data reveals a $2\sigma$ indication of dynamical dark energy evolution at $z<1$, 
featuring a phantom crossing at $z \simeq 0.4$, in agreement with the DESI DR2 findings -- 
although the deviation from $w = -1$ at lower redshift is less pronounced than that reported by DESI.
The overall trend of the reconstructed $w(z)$ shows a decrease with increasing redshift. 
However, the confidence level diminishes at higher redshifts due to larger uncertainties, 
stemming from the limited availability of observational data in that regime.

To reconstruct the interaction between dark energy and dark matter, 
the equation of state of dark energy must be provided \textit{a priori} to break the degeneracy.
Instead of directly reconstructing $f$ and $\dot{f}$, 
we reconstruct $\Omega_f = \Omega_c \frac{f}{f_0} + \Omega_b$ and its derivative, 
thereby avoiding potential issues related to the priors of $f$, $\Omega_c$, and $\Omega_b$ at z = 0. 
Our focus is on whether $\dot\Omega_f=0$, which is independent of the specific values of $f_0$, $\Omega_c$, and $\Omega_b$. 

In this paper, we adopt two priors for the equation of state by extracting constraints on the CPL parameterization of $w(z)$ from both Planck~\citep{Planck:2018vyg} and DESI DR2~\citep{DESI:2025zgx}. 
The Planck results yield $w_0 = -0.957 \pm 0.080$, $w_a = -0.29^{+0.32}_{-0.26}$, consistent with the $\Lambda$CDM model, while the DESI DR2 results provide $w_0 = -0.838 \pm 0.055$, $w_a = -0.62^{+0.22}_{-0.19}$, indicating a significant signal of dynamical dark energy. 
The covariance between $w_0$ and $w_a$ is included in both cases.

\begin{figure*}
\includegraphics[width=0.45\textwidth]{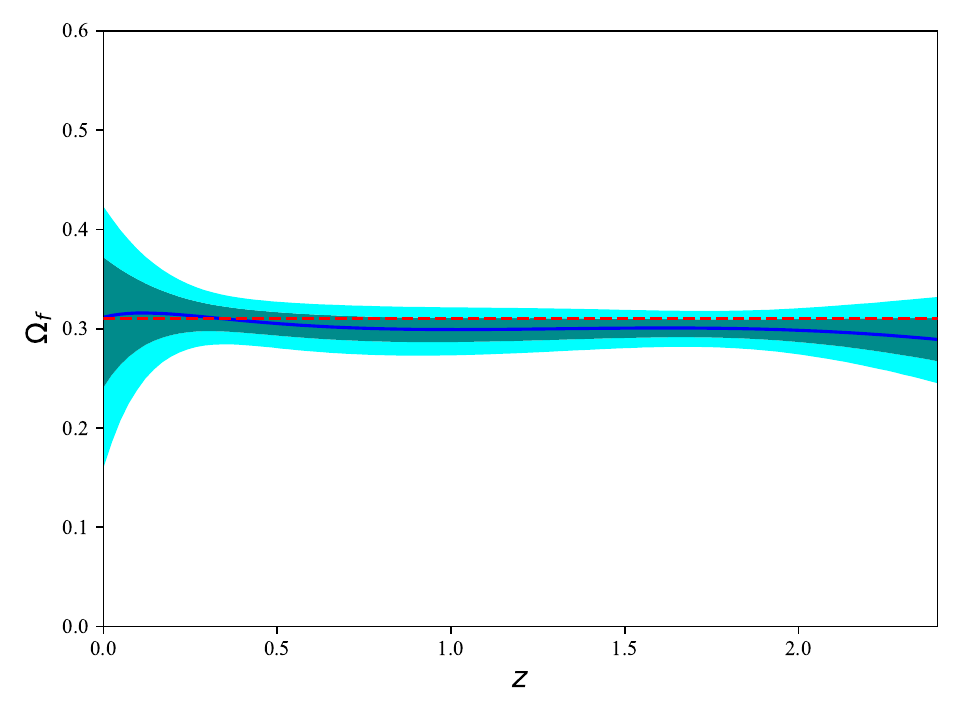} 
\includegraphics[width=0.45\textwidth]{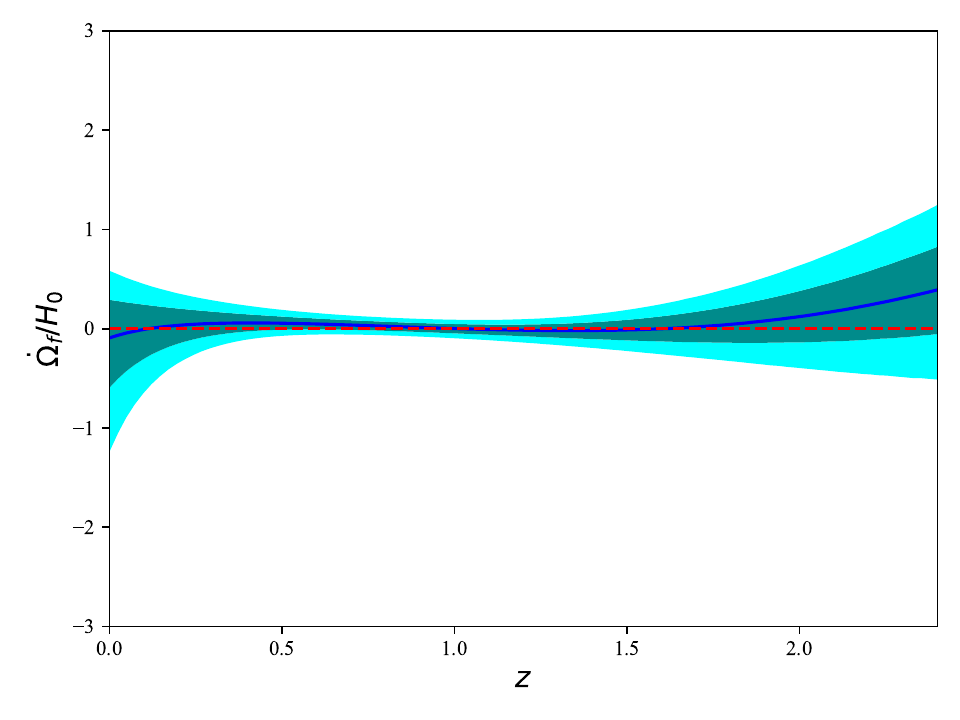} 
\includegraphics[width=0.45\textwidth]{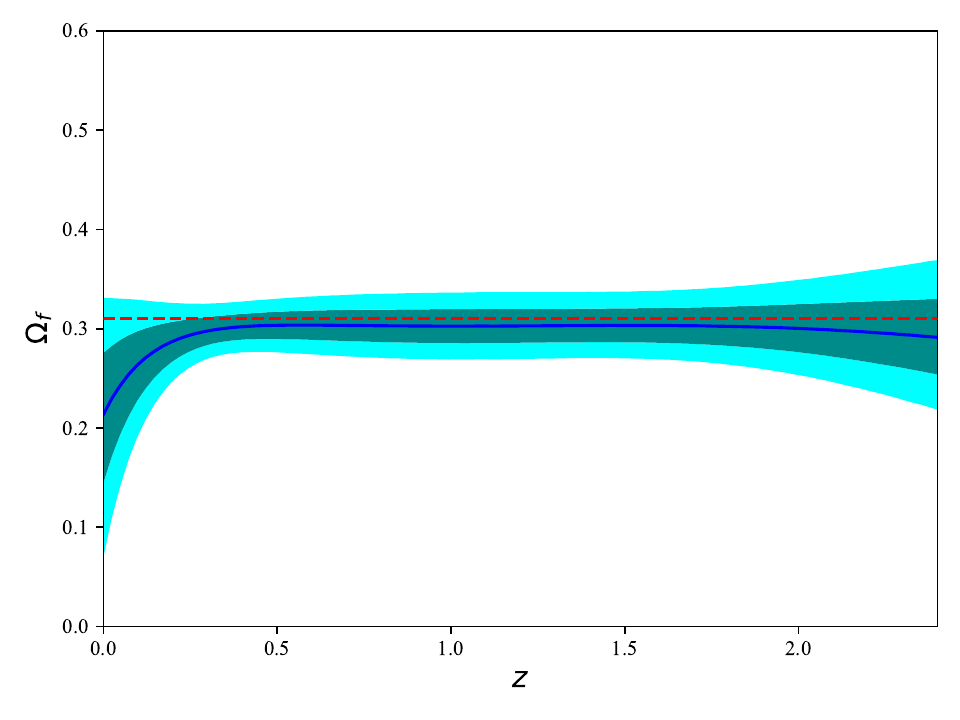} 
\includegraphics[width=0.45\textwidth]{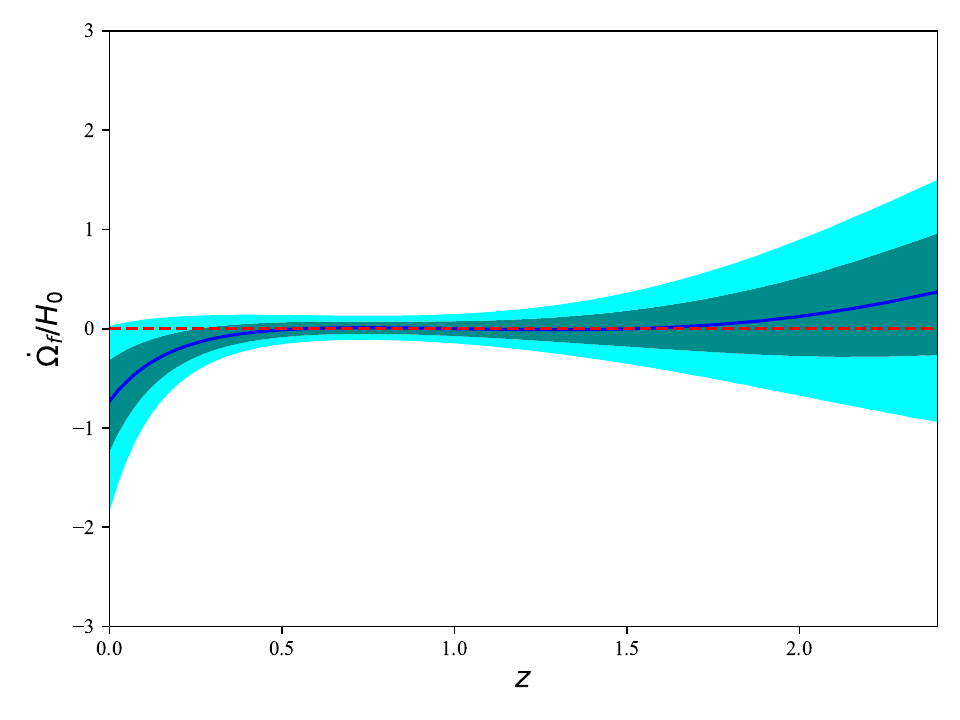} 
\caption{Reconstructions of the dark sector interactions from DESI DR2+CC+OHD+SNe Ia. We adopt two priors of $w(z)=w_0+w_a\frac{z}{1+z}$ which are extracted from Planck (upper panel) and DESI DR2 (lower panel). The dashed lines denote $\Omega_m=0.31$ from Planck (left panel) and zero coupling (right panel), respectively.}
\label{fig:coupled}
\end{figure*}

As shown in the upper panel of Fig.~\ref{fig:coupled}, 
the reconstruction of dark sector interactions -- based on the combined datasets from DESI DR2, OHD, CC, and SNIa using Planck priors --
is consistent with a non-coupled dark energy scenario.
The reconstructed $\Omega_f$ is also consistent with the value of $\Omega_m$ derived from Planck.
Compared to previous work~\citep{Yang:2020jze}, the previously reported $>1\sigma$ coupling at high redshift is now reduced to within $1\sigma$.

Interestingly, as demonstrated in the lower panel of Fig.~\ref{fig:coupled}, when the dynamical EoS prior from DESI is adopted, a $\sim 2\sigma$ signal of dark sector interactions emerges at $z < 0.5$. 
A negative $\dot\Omega_f$ at low redshift indicates energy transfer from dark matter to dark energy, consistent with results obtained using a parametric approach in~\cite{Giare:2024smz}. 
We emphasize that our analysis is independent of any specific cosmological model or assumed interaction form.
Our results suggest that, from a model-independent and data-driven perspective, 
dark sector interactions will inevitably emerge if the dynamical dark energy indicated by DESI is confirmed.

\section{Conclusion and discussions}\label{sec:conclusion}
In this paper, we present a data-driven, model-independent reconstruction of the dark energy equation of state and its interaction with dark matter using cosmic expansion data.
By incorporating the new DESI DR2 BAO measurements into the previous CC+OHD+SNe Ia combination, 
we identify a $2\sigma$ dynamical signature in the uncoupled dark energy scenario, 
including a phantom crossing around $z \sim 0.4$, consistent with DESI’s findings.
Intriguingly, when the constraints on the CPL form of $w(z)$ from DESI are adopted as a prior to reconstruct the interaction between dark energy and dark matter, a $\sim 2\sigma$ signal of non-zero coupling at low redshift emerges.
Our findings suggest that the significant dynamical signal of dark energy revealed by recent DESI data may also point to a notable interaction between the dark sectors. 
We need to investigate both the dynamics and interactions of dark energy as more data and improved methodologies become available in the future.

Employing a data-driven, model-independent GP framework, we determine the fiducial value of $H_0$ and reconstruct $E(z)$ along with its derivatives. 
While the reconstruction of $H_0$ is sensitive to both dataset selection and the choice of GP kernel (as discussed in~\citet{Johnson:2025blf}), 
we adopt the Matérn-9/2 kernel to mitigate potential overfitting artifacts associated with the widely used Radial Basis Function (RBF) kernel,
which can amplify observational uncertainties~\citep{Johnson:2025blf}. 
The RBF kernel yields $h = 0.6932$ (where $H_0=100h~\mathrm{km\,s^{-1}\,Mpc^{-1}}$), slightly higher than the fiducial value $h = 0.6887$ used in our analysis. 
However, dependencies on $H_0$ cancel out in the subsequent reconstructions of $w$, $\Omega_f$, and $\dot{\Omega}_f/H_0$, 
rendering our results robust to the choice of kernel. 

Similar or alternative reconstruction techniques have been employed to explore various cosmological problems~\cite{Gomez-Valent:2018hwc,Cai:2019bdh,Belgacem:2019zzu,Benisty:2020kdt,Bernardo:2021qhu,Benisty:2022psx}.
For example,~\citet{Benisty:2020kdt} used GP to reconstruct the matter growth history and examine the $S_8$ tension using Redshift Space Distortion data.
\citet{Benisty:2022psx} applied both GP and Artificial Neural Networks (ANN) to assess the stability of the Supernova absolute magnitude.
\citet{Gomez-Valent:2018hwc} used GP to estimate the Hubble constant, while~\citet{Belgacem:2019zzu} employed GP to constrain modified gravity theories using gravitational wave observations.
The core methodology in these studies aligns with the approach adopted in this work.
These results illustrate that non-parametric reconstruction techniques, in combination with flexible kernel functions, can effectively capture key features in cosmological data sets.

Our methodology is inherently flexible within the data-driven paradigm, 
accommodating a wide range of machine learning (ML) techniques for regression and parameter reconstruction from large cosmological datasets. 
While GP is commonly used for non-parametric reconstructions, our framework is readily extendable to other ML approaches, 
including Neural Networks (NNs)~\citep{Fluri:2018hoy} and Deep Learning~\citep{George:2017pmj}. 
This versatility ensures that our approach remains adaptable to a wide range of problems and applicable to additional cosmological data sets (e.g., cosmological perturbation data). We leave such extensions for future work.

\begin{acknowledgments}
This work is supported by ``the Fundamental Research Funds for the Central Universities'' under the reference No. 2042024FG0009. The numerical calculations in this paper have been done on the supercomputing system in the Supercomputing Center of Wuhan University. 

\end{acknowledgments}

\bibliography{ref}

\end{document}